\documentclass[10pt,a4paper]{article}
\usepackage{amsmath}
\usepackage{amsthm}
\numberwithin{equation}{section}
\usepackage{graphicx}
\title{  
\textsc{Ranking structures and Rank-Rank Correlations  of  Countries. \\The FIFA and UEFA cases}
}
\author{ Marcel Ausloos$^{1,}$\footnote{Associate Researcher at eHumanities group,
Royal Netherlands Academy of Arts and Sciences, Joan Muyskenweg 25, 1096 CJ Amsterdam, The Netherlands. } , Rudi Cloots$^2$, Adam Gadomski$^3$,   Nikolay K. Vitanov$^4$  }
\date{ 
$^1$ GRAPES,  rue de la Belle Jardiniere, 483/0021\\
B-4031, Liege Angleur, Euroland  \\ email: marcel.ausloos@ulg.ac.be\\   $^2$  University of  Li\`ege, \\ Department of Chemistry, B6C LCIS - GreenMAT,  B-4000 Li\`ege\\email:  rcloots@ulg.ac.be \\   $^3$  University of Technology and Life Sciences, \\ Department of Physics, Institute of Mathematics and Physics, PL-85-796 Bydgoszcz, Poland\\email:  agad@utp.edu.pl;agad@atr.bydgoszcz.pl   \\$^4$ Bulgarian Academy of Sciences, Institute of Mechanics, \\
Acad. G. Bonchev Str., Bl. 4, BG-1113 Sofia, Bulgaria\\email: vitanov@imbm.bas.bg}
\begin{document}
\maketitle
\begin{abstract}
  Ranking   of agents  competing with each other in complex systems  may  lead to paradoxes  according to  the pre-chosen different measures. A discussion is  presented  on such rank-rank, similar or not, correlations based on the case of  European countries ranked by UEFA and FIFA from  different soccer  competitions.   The first question to be  answered  is  whether an empirical and simple law  is obtained for such  (self-) organizations of complex sociological systems
   with such different measuring schemes. It is found that the power law form is not the best description contrary to many modern expectations. The stretched exponential is much more adequate. Moreover,  it is found that the measuring rules lead to some inner structures, in both cases.
\end{abstract}

\section{Introduction }\label{introduction}.

It is of common knowledge that in life one has often to choose some product. Similar products are ranked, usually according to  some criterion.  Several criteria can be  employed for influencing one's choice. However the criteria may lead to different ranking lists, albeit the bare quality of the product
should rationally be taken as criterion-independent. In several cases, this leads to incompatibility. One famous example, known as Arrow's theorem \cite{[AR]}  implies that a choice might not be always logically possible.  Practical cases for example occur  in sports, like college football in the NCAA, where the best teams are ranked according to  {\it  voting} procedures,  by different media.  Another  team sport, soccer, is worldly known as also enticing enthusiasm and discussions.  Teams  (clubs or countries) are ranked  through results of various competitions. However round robin tournaments, to decide the best team,  are rare, although in theory  are  the fairest ways to determine a champion among a given number of participants.   Of course, a round robin tournament can also be used to determine which teams are the poorest performers.

It is well known that a knockout tournament where half of the participants are eliminated after each round is a faster method of selection, but  the method is highly debatable since it matches teams (or players) somewhat randomly. The final results being therefore often dependent of the draw, i.e. with biased initial conditions.

Finally, the so called Swiss system tournaments attempt to combine elements of the round-robin and elimination formats. This is very usual:  the ranking of  soccer countries, studied here, belongs to such a category .

Note that the same considerations can be made for players in individual sports  and also when players are grouped by pairs or triplets, e.g., in golf,  tennis or billiards,  petanque (french game of bowls), or more.
\par

The specific scientific  literature on  soccer  $ ranking$ themes  seems limited to (i)   a 2001  paper  by Kern and   Paulusma   \cite{DAM108.01.317} who discussed FIFA rules  complexity  for   competition outcomes leading to ranking and (ii) a
  2007 paper by Macmillan and  Smith, explaining $country$  ranking  \cite{JSE08.07.202explainingrankingsoccercountries}.  From a more general point of view, one should mention 
Churilov and Flitman  \cite{COR33.06.2057} 
  model for producing a ranking of participating teams  or countries, like in  olympics  games.
    
\par
 
There are other  papers, quite interesting,  - since at the interfaces of various disciplines,   often tied to various technical questions or  limited to the analysis of distribution functions,  - thus without conveying questions  on e.g.  (self-) organizations of complex sociological systems.

 From a scientific point of view, ranking is an old problem with a
long history.    The  comparison of    "values", through ranking,   has produced dozens of ranking methods,  not only in sports \cite{JAS24.97.635worldsportsratingStefani}, but also for  ranking
 candidates in a political  context, scientists, webpages or various types of "goods" or "agents".   
  Among many nonparametric procedures relying on counting and ranking processes applied either directly to the sample data or to some natural function,   the most basic, but immediately useful, analysis pertains to ideas following Zipf ranking considerations \cite{z1}.  Thereafter several empirical laws can be imagined as recalled in Sect. \ref{rankinglaws}.

However, measurements or ranking in sport competitions, though frequently reported in the media,  often lack the necessary descriptive power, as the physics of complex systems  usually present,  - with a recent  exception \cite{[35]}. Here below,  an analysis of   some ranking data from a specific nonlinear complex system, i.e.  soccer  country ranking,  as a specific modern society interesting example, is reported. Two   systems, i.e. the FIFA  and the UEFA "measures" are explained and illustrated in Sect. \ref{countryranking}.
The  section presents  a short description of the studied data and of the ranking rules.  The rank-size relationship is searched for the   FIFA or UEFA ranking, respectively.  At first, as in most modern studies on ranking in complex systems of  interacting agents, power laws are  expected. It will be found that they are $not$ subsequently observed in  neither cases.  
A general discussion of the displayed features  and specific  comments, leading to some understanding of the findings, are presented.  

The article ends with a discussion of rank-rank correlations as can be  proposed from these  two sets of data, - and the measuring criteria, in Sect.\ref{rankrank}, before   a brief pragmatic discussion of findings in Sect. \ref{discussion} and conclusions in Sect.\ref{conclusions}. Two Appendices are also attached to this article.  In  Appendix A, the ranking of $all$ FIFA countries  is briefly analyzed  for completeness. Another (Appendix B)  shows the  unexpectedly simple, but yet not  explained,  "trivial" correlation between two UEFA ranking coefficients.

\section{On Ranking Laws}\label{rankinglaws}

Ranking analysis has  been performed for a long time since Zipf \cite{z1} who  observed  that a large number 
of  size  distributions, $N_r$ can  be approximated by a simple  {\it 
scaling (power) law} $N_r = N_1/r $,   where  $r\ge1$ is the ranking parameter, with 
 $N_{r } \ge N_{r+1}$, (and obviously  $r<r+1$). Many developments followed such an observation,   e.g.  of course in  linguistics,  
 but also in many domains \cite{gabaix}. Whence,   Zipf ideas has led to a flurry of log-log diagrams showing a straight line through the displayed data. It is obvious that a more flexible equation/law,
\begin{equation}\label{Zipfeq}
y_r = \cfrac{a}{r^\alpha},
\end{equation}
is of greater interest,  since it has two parameters; it is called the {\it rank-size scaling law}. The particular case  $\alpha=1$ is thought to 
represent a desirable situation, in which   forces of concentration
balance those of decentralization \cite{gabaix}. Such a case is called  the rank-size $rule$ \cite{gabaix}-\cite{PhA391.12.767ranksizescaling}. 
 The scaling exponent can be used to judge whether or not the size distribution is close 
to  some optimum  (equilibrium) state.

The rank-size relationship   has  been frequently identified and sufficiently discussed 
to allow us  to base much of the present investigation on such a simple law.  
This may
be "simply" because the rank-size relationship can be reached 
from a wide range of specific situation  \cite{[34]}-\cite{PhA391.12.767ranksizescaling}.  
\par

However, the mere exponential  (2 parameter fit) case  \begin{equation} \label{PWL}
 y(r)=  \;b\;e^{-\beta r}
\end{equation} should be also  considered. For short hand notations, these two analytic forms are called Pwl2 and Exp2   in the figures.

 In the present paper,    three other  often used  3-parameter  statistical distributions,  generalizing the  power  and/or exponential law   are used to examine the   UEFA  data:  (i) the  Zipf-Mandelbrot-Pareto  (ZMP) law reads
 \begin{equation} \label{ZMeq3}
y(r)=\hat{c}/(\eta+r)^{\zeta} \;\equiv \; [c/(\eta+r)]^{\zeta},
\end{equation}

(ii) the power law with cut-off reads
  \begin{equation} \label{PWLwithcutoff}
 y(r)= d \;r^{-\gamma} \; e^{-\lambda r},
\end{equation}

(iii) while  the stretched exponential is
 \begin{equation} \label{Stxeq3}
y(r)=\theta \; x^{\mu-1} \; e^{-\nu\;x^\mu}.
\end{equation}

 For short hand notations, these three analytic forms are called ZMP3,  Pwco3, and Stx3, respectively,  in the figures. Since $\eta$, in  Eq.(\ref{ZMeq3}),  is not necessarily found to be an integer in a fit procedure, $r$    can be considered  as a continuous variable,  for mathematical convenience, without any loss of mathematical  rigor; the same for the fit parameters $a$, $b$, $c$ (or $\hat{c}$),  $d$,  and $\theta$, and  for  the "relaxation ranks" $\beta$ and $\lambda$ (and $\nu$).   
Note that both $\alpha$ and $ \zeta$ exponents, in  Eq.(\ref{Zipfeq}), and  Eq.(\ref{ZMeq3}). must be greater than 1 for the distributions to be well-defined (also greater than 2 for the mean to be finite, and  greater than 3 for the variance to be finite). 

   \begin{figure}
\centering
 \includegraphics [height=16.5cm,width=14.5cm]{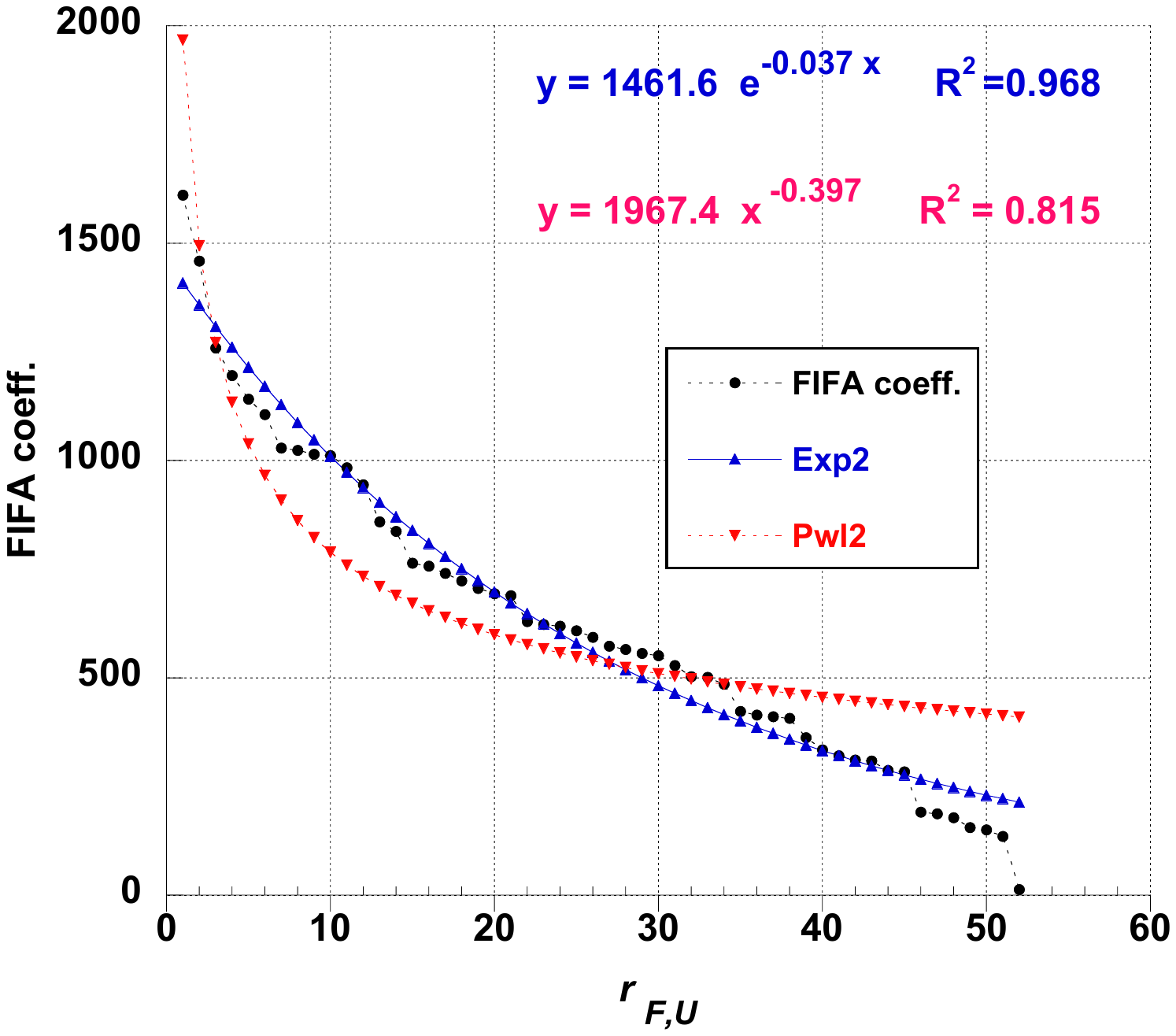}
 \caption   {    An exponential and a power law fit are shown,  with the numerical  values of parameter fits and  the corresponding regression coefficient $R^2$, suggesting possible and simple empirical relationships between  the  FIFA coefficient for  the  ranked, in descending order, 53  UEFA "Association Members"    in Sept. 2012 }   \label{Fig1Plot10_2fitslili} 
\end{figure}

   \begin{figure}
\centering
 \includegraphics [height=16.5cm,width=14.5cm]{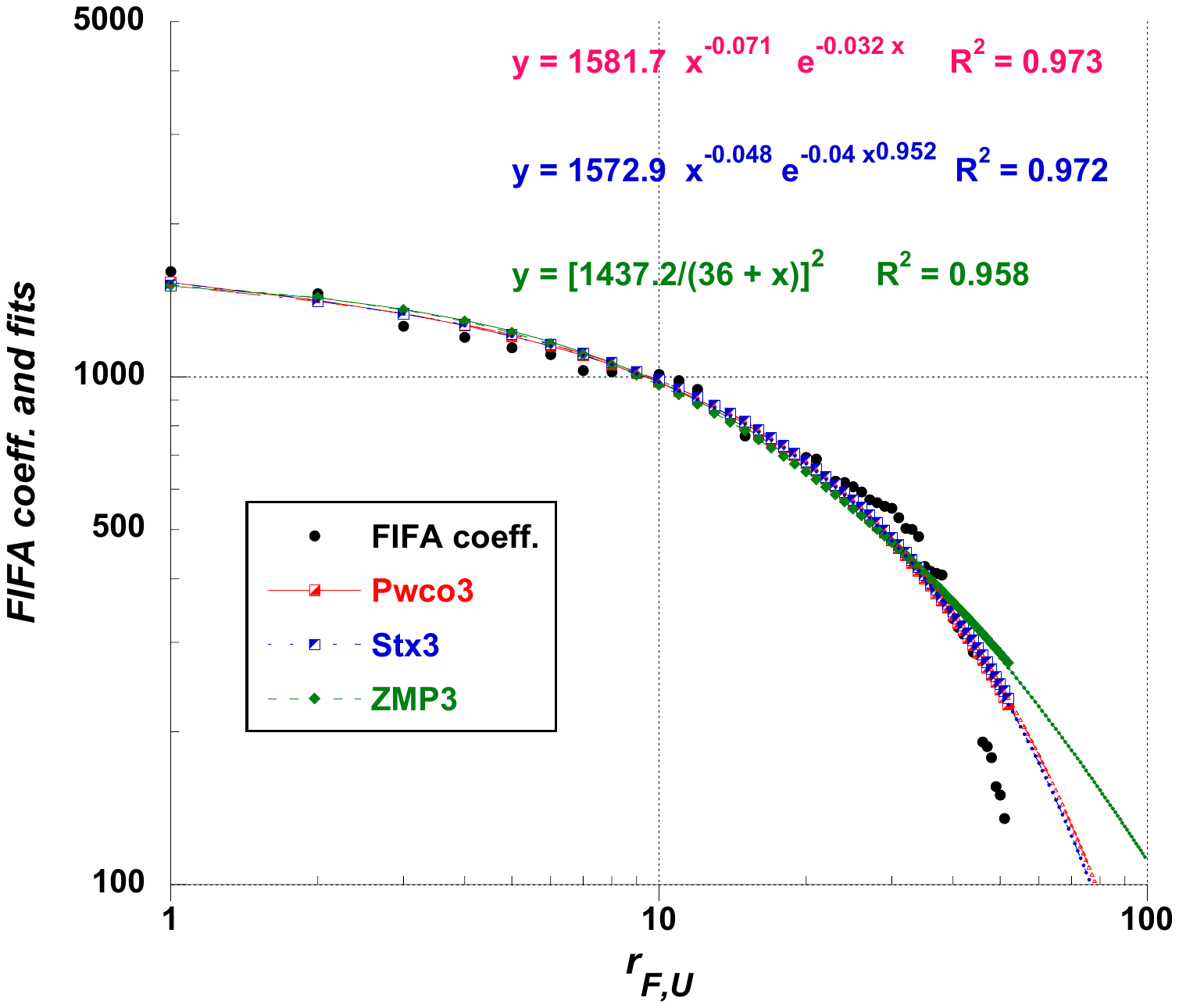}
\caption   {  Possible empirical relationships between  the  FIFA coefficient for  the 53  UEFA "Association Members"  (ranked in descending importance order) in Sept. 2012; three 3-parameter law fits are shown on a log-log plot.  The   last  3 data points at high rank, with FIFA coefficient less than 100, are not displayed for enhancing clarity. The   corresponding regression coefficient $R^2$  are also given  in Table 2  for comparison with other fits} 
 \label{Fig3Plot10_3fitslolosh} 
\end{figure}

     \begin{figure}
\centering
 \includegraphics [height=16.5cm,width=14.5cm]{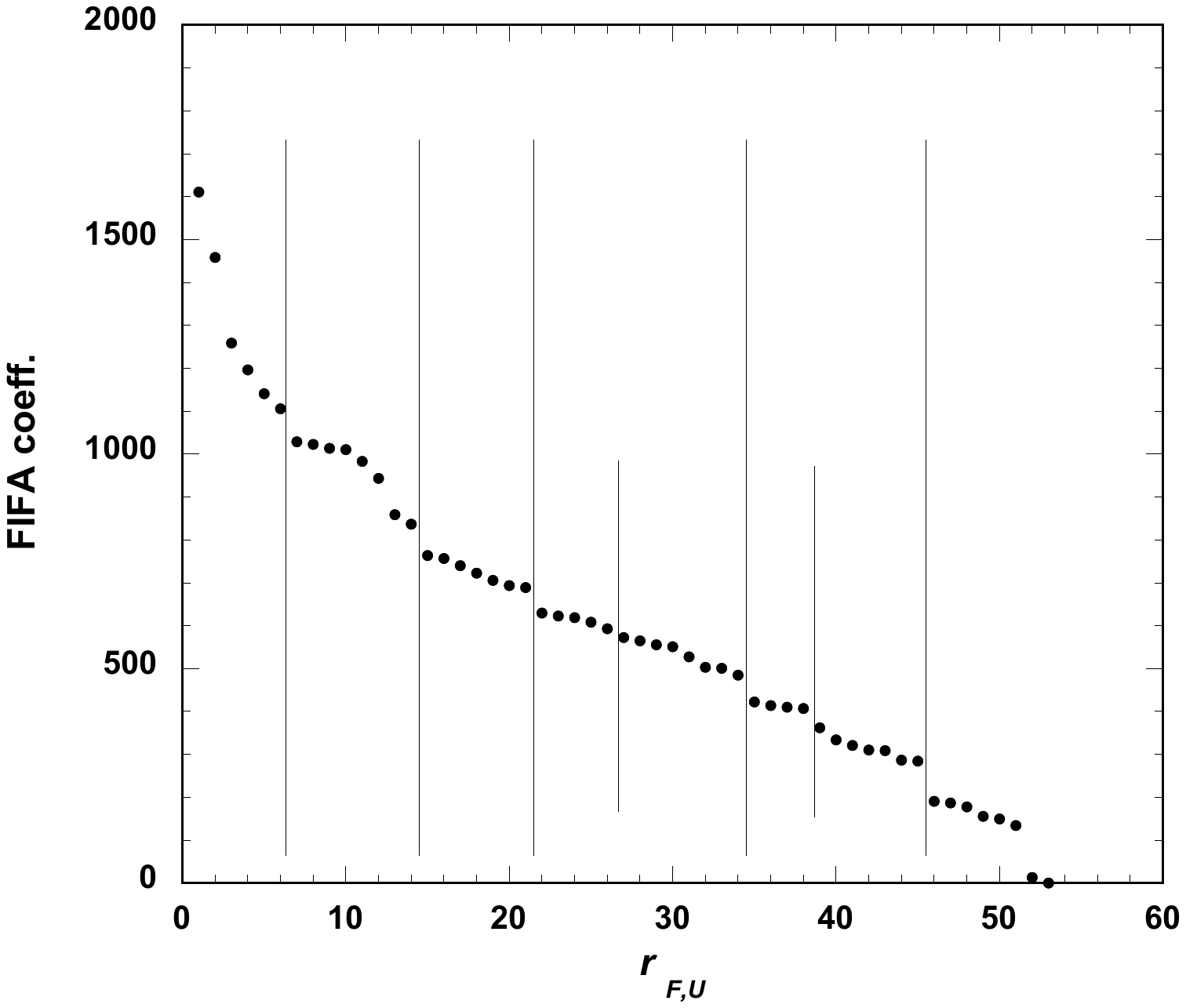}
\caption   { FIFA coefficient for  the 53 UEFA "Association Members"  ($ \sim$ $countries$)  as ranked in decreasing order  in Sept. 2012.  "Regimes" are pointed out by vertical lines.    The average regime size $\simeq 6.7$ with a (high)  $R^2$= 0.994 }
 \label{Fig4Plot3FIFA0912regUEFAlili} 
\end{figure} 
 
\begin{table} \begin{center} 
\begin{tabular}[t]{cccccccccc} 
  \hline 
   $ $   &FIFA &UEFA  &UEFA& FIFA    \\ 
   $ $   &(UEFA)   &"mean"  &"points" & all    \\ 	
\hline  
min&  0 &	0&0.92&   0   \\
Max&	 1617&20.857&84.41&1617  \\
Sum&32647&234.41&1098.5&81830\\
mean ($\mu$)&	615.98&4.4229&20.726&391.53 \\
median (m) &	606&2.3&14.25&328\\
RMS & 711.23&6.460&29.75& 501.17 \\
Std. Dev. ($\sigma$) & 358.95&4.753&21.553&313.6\\
Var. &128846.7&22.59&464.5&98346.1 \\
Std. Err. &	49.306&0.65287&2.960&21.692  \\
Skewn.  &0.6268&1.470&1.486&1.065 \\
Kurt. &0.054&1.646&1.517&1.069\\ \hline
 $\mu/\sigma$ &1.716    &0.930&0.962 &1.248&   \\
  3\;$(\mu-m)/\sigma$ &0.0834&1.340&0.901 &0.608&   \\ \hline
  $N$&	 53&	53& 53 & 209 \\
$N_0$ &	1 &	2&0&4 \\ \hline
\end{tabular} 
   \caption{Summary of  statistical  characteristics  for  Sept. 2012 of national teams ranking data according to different "measures"; $N$ is the number of data points; $N_0$ the number of  teams with a 0 value coefficient    }\label{Tablestat}
\end{center} \end{table} 

  \section{Country ranking }\label{countryranking}

  First, let   the major difference in the ranking of national  squads  by FIFA  \cite{FIFArankingrules} and  the ranking  of  countries\footnote{More exactly called "Association Members": often each  "agent" represents one country, though not  necessarily: e.g., England, Scotland, Wales, Northern Ireland, Faroe Islands, New Caledonia, etc. are distinguished.}  by UEFA  \cite{UEFAcountryrankingrules} be emphasized. The former results from ranking following a set of matches between national teams,   the latter is  deduced from  team ranking of the  Association Members  \cite{UEFAteamrankingrules}. The former serves to allocate primacy in drawing of groups at the World Cup, the latter to calculate how many teams per "country" are allowed to play in the Champions League and in the Europa League. Both rankings  serve  {\it a priori} different  competitions and aims, both tied to different economic conditions or interests, -  a discussion outside our present  purposes.  However, the fact that such rankings  are for the same "objects" or "agents" is the underlying signature of the present  scientific questions, mentioned in the Sect.\ref{introduction}.
  
  In Sept 2012, FIFA  \cite{FIFA}, made of  6 confederations \cite{Confacronym}, grouped  209 Member Associations squads,  $\sim"countries"$,   53 of them being in  the UEFA.   Therefore the FIFA  "country" rank value, $r_F$, has to be "projected" into the UEFA rank space, $r_{F,U}$, eliminating out-of-UEFA squads, - thereby giving an appropriate new rank value for subsequent data analysis, and for comparison with the UEFA rank, $r_U$.

  In the following there is no discussion of the evolution of the rank of  any country; the data pertains to both ranking exclusively in Sept 2012.  Nevertheless, it is expected that the  chosen date, relatively  arbitrary, away from major events,  can serve as a valid one for the questions and discussions pertaining to such a type of data set .

        \subsection{ FIFA Country ranking }\label{FIFAcountryranking}
        
The FIFA Country ranking system  is  based on results over the previous four years (instead of the previous eight years) since July 2006. It is briefly described for completeness in Box\#1.

 \vskip0.5cm
\fbox{%
\begin{minipage}{5 in}
Box $\#1$
 \vskip0.2cm 
{\bf FIFA points system}
\vskip0.5cm

The total number of points  depends on results over a $four$ $
 year$ period as determined by adding:
\begin{itemize} \item the $average$ number of points gained from matches during the past 12 months  
  \item  and the $average$ number of points gained from matches older than 12 months, thus over the 3 previous years 
  \end{itemize} 

The number of points that can be won in a match depends on (*)  
\begin{itemize} \item  the match result (win, draw, loss) ($M$)
\item  the match importance ($I$)
\item  the ranking of the opposite  country ($T$),
and \item the confederation to which the opponent belongs ($C$)
  \end{itemize} 
 The total number of points  ($P$) is found from
\begin{equation}\label{FIFAcount} 
P = M \;.\; I \;.\;  T \;.\;  C 
\end{equation}
(*) In short,
\begin{itemize} \item  $M$=3, 1, 0   for a win, draw, loss
\item   $I$= 1.0, 2.5, 3.0, 4.0 respectively for any friendly, world cup qualification,  confederation level final competition,   world cup final competition match
\item    $T$ = 200 - the ranking position of the opponent\footnote{ 
As an exception to this formula, the team at the top of the ranking is always assigned the value 200 and the teams ranked 150-th and below are assigned
an equal   value of 50.}.
The ranking position is taken from the opponent ranking  in the most recent rank list published by FIFA, 
and \item  $C$= 1.0 for UEFA and CONMEBOL, 0.88 for CONACAF, 0.86 for AFC and CAF, and 0.85 for OFC (see \cite{Confacronym}  for acronyms)
  \end{itemize} 
  
\end{minipage}}

\vskip 0.5cm
\subsection{FIFA(UEFA) ranking data analysis}\label{FIFAdatanalysis}

First, the UEFA countries, as ranked by FIFA,  are first extracted from the overall rank distribution, and their order $r_F$ reconstructed such that there is no gap in the ranking, $r_{F,U}$. The statistical characteristics of this "projected" rank distribution are given in Table 1.

Next, the simple exponential and the power law as fitted to the  FIFA coefficient for  the  ranked, in descending order, 53  UEFA "Association Members"    in Sept. 2012   are shown in Fig.\ref{Fig1Plot10_2fitslili}. The numerical  values of parameter fits and  the corresponding regression coefficient $R^2$ are given.  It is at once remarkable that the power law fit is a poor approximation to   fit the data.
  
   Possible other empirical relationships between  the  FIFA coefficient for  the 53  UEFA "Association Members"  (ranked in descending importance order) in Sept. 2012  are shown on a log-log plot in Fig.\ref{Fig3Plot10_3fitslolosh}.  The   last  3 data points, at high rank, with FIFA coefficient less than 100, are not displayed for enhancing clarity. The three 3-parameter law fits are astoundingly similar and rather undistinguishable. However, the ZMP3 fit, i.e. the more simple generalizing the power law, appears to be the worse of the three fits. The   corresponding regression coefficient $R^2$  are also given  in Table 2  for comparison with other fits.  
   
   From  the classical linear-linear axes plot in Fig.\ref{Fig1Plot10_2fitslili} and also from the log-log plot in Fig.\ref{Fig3Plot10_3fitslolosh}, some  deviation can be observed. However there are not irregularly distributed on both sides of the fits, but appear as "regimes", through jumps at specific data points, apparently rather regularly spaced. 
   
   The UEFA distribution coefficient characteristics  are given in Table 1.  The kurtosis and skewness  suggest  the existence of "structures". 
 They are pointed out in  Fig. 
 \ref{Fig4Plot3FIFA0912regUEFAlili},  displaying the Sept. 2012 FIFA coefficient for  the 53 UEFA "Association Members"  ($ \sim$ $countries$)  ranked in decreasing order.  The "regimes" are pointed out by vertical lines.    The average regime size is remarkably well defined, i.e. $\simeq 6.7$ with a  $R^2$= 0.994. 
 
 {\it A posteriori}, it might have been expected that a regime size = 8 might have been underlying, since 8 = 2$^3$ (!), which is, as other values of $2^m$, usual integers for defining groups and pools in various tournament competitions. The value 6.7 seems to indicate that the intrinsic border between groups is fluctuating, i.e. only 1 or 2 teams can move from one group to another.

     \begin{figure}
\centering
 \includegraphics [height=16.5cm,width=14.5cm] 
 {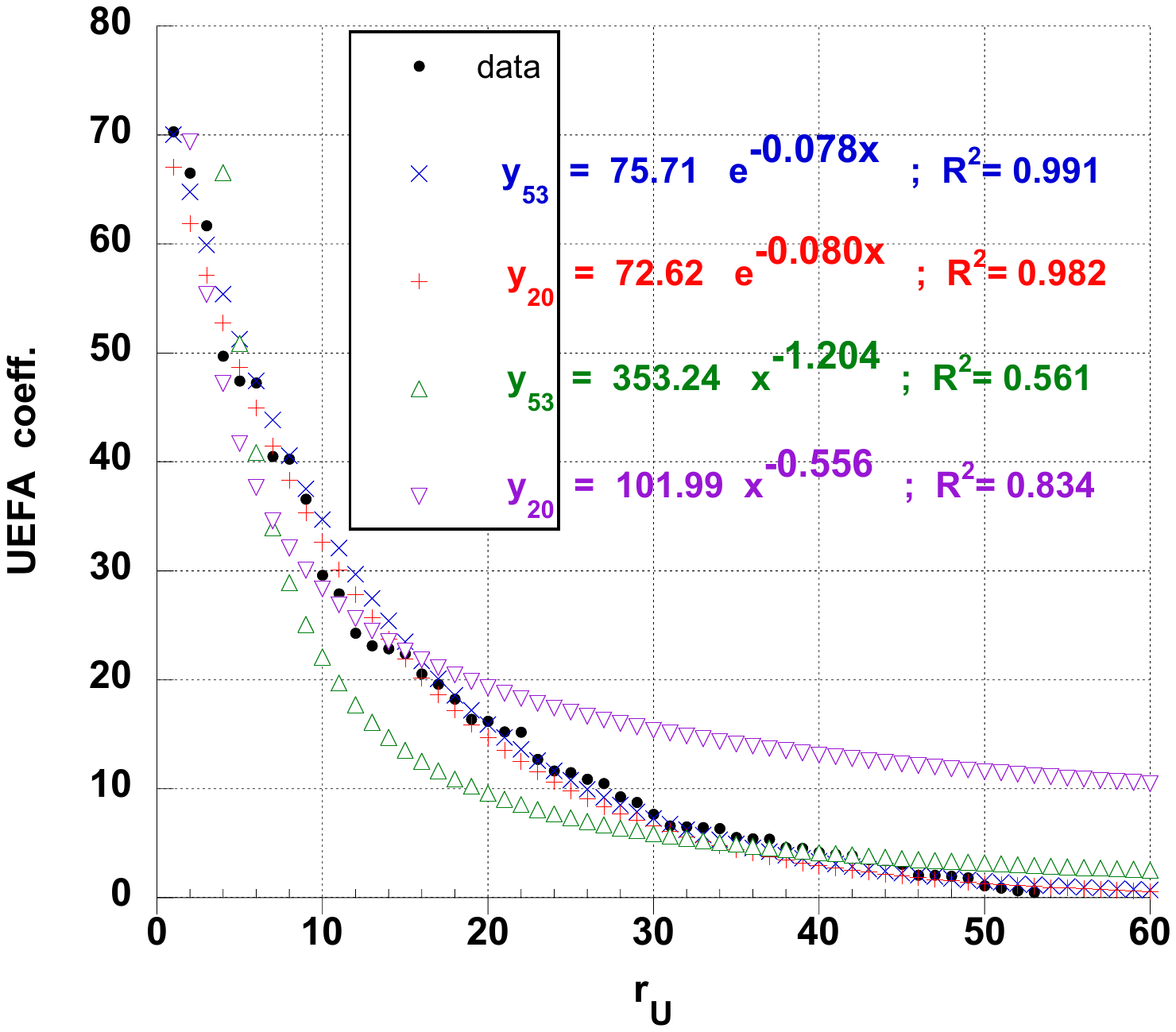}
 \caption   {  Possible simple empirical laws of the UEFA coefficient  (i) for  the 53 "Association Members" ( $\sim countries$)  and  (ii) only for those $below$  the  20 ranking, and their rank  in Sept. 2012. In each case, both an exponential and a power law fit are  given with the numerical  values of parameter fits and  the corresponding regression coefficient $R^2$  } 
 \label{Fig6Plot8UEFA4yrnkcountries} 
\end{figure}

 \begin{figure}
\centering
 \includegraphics [height=17.5cm,width=14.5cm]{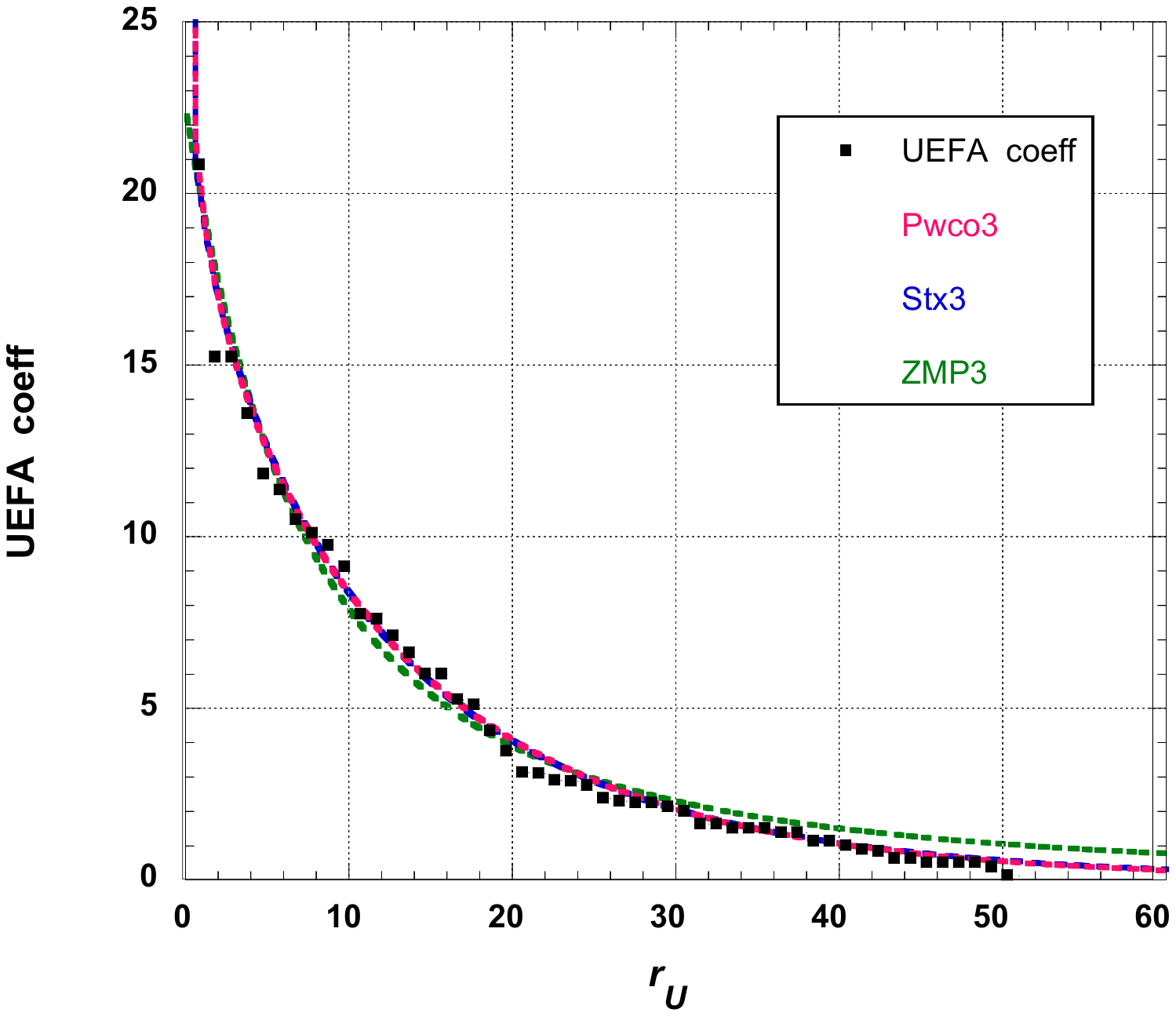}
\caption   {   Empirical relationships between  the  UEFA coefficient ($U_c$)  for  the 53 "Association Members" ( $\sim countries$),   in decreasing order, and their rank  ($r_U$)  in Sept. 2012, for  three 3-parameter law fits;  the corresponding regression coefficient $R^2$  are given in Table 2 for comparison with other fits } 
 \label{Fig5Plot36UEFA1112coco3flili} 
\end{figure}

  \bigskip
    \subsection{ UEFA Country ranking }\label{UEFAcountryranking}

The UEFA (association member or) $country$ ranking takes into account the results of all $clubs$ from each association.  (It is used to determine the number of entries an association is granted for forthcoming seasons.)    This UEFA club coefficient ranking is  based on the results of all European clubs in UEFA club competition, i.e. the  UEFA Champions League and  the UEFA Europa League, on the $five$ previous "seasons".    The calculation of the "country" coefficient is recalled in Box\#2.  The  club coefficients of success is described on   \cite{UEFAteamrankingrules} and reworded for completeness in Box $\#2$ as well. Observe that the rules are more complicated than  a "win-draw-loss"  rating.

Two UEFA  country measures exist  \cite{UEFAcountryrankingrules}:  the  point ranking $U_p$ and the  mean ranking $U_c$. A country rank, either $r_{Up}$  or  $r_{Uc}$ slightly differs according to the measure; see Appendix B. The  mean ranking is used here below for analysis.  The statistical characteristics of both distributions are given in Table 1.

  \vskip0.5cm
\fbox{%
\begin{minipage}{5 in}

Box $\#2$ \vskip0.2cm 

 {\bf UEFA country ranking systems}
\vskip0.5cm
 
The total number of points   results from the sum of  the number of points  teams of an Association have won over a $five$ $ years$ period.

\begin{itemize} \item  The coefficient  $U_c$ is calculated by working out an average score: dividing the number of points $U_p$ 
(*)
 obtained, by the total number  $N$ of clubs  having represented an association in  both UEFA club competitions., within that season.
\item The resulting figure is then tallied with the results of the previous  $four$ seasons to calculate the UEFA  $ country$ coefficient.  
  \end{itemize}

(*) In short, the number of points  won  by  a team of an Association  is calculated as follows:

\begin{itemize} \item A team gets 2 points for a win and 1  point for a draw, but \item points are halved for matches in the qualifying and play-off rounds.
\item   4 points are awarded for participation in the group stage of the UEFA Champions League and 4 points for qualifying for the round of 16.
\item  Clubs that reach the round of 16, quarter-finals, semi-finals or final of the UEFA Champions League, or the quarter-finals, semi-finals or final of the UEFA Europa League, are awarded  1 extra point, {\it  for each round}. 
  \end{itemize}  
  
     \vskip0.2cm  The UEFA rankings are updated after each round of UEFA club competition matches.
\end{minipage}}

   \vskip 0.5cm
\subsection{UEFA ranking data analysis}\label{UEFAdatanalysis}

As above, several analytical expressions have been used,
i.e. starting from an exponential law and a simple power law, and their generalization to a 3-parameter analytical form for searching an interesting empirical  relationship  between  the  UEFA coefficient for  the 53 "Association Members" ( $\sim countries$)   and their rank  in Sept. 2012,  both the exponential and the power law fit, shown in Fig.\ref{Fig6Plot8UEFA4yrnkcountries}, are hardly convincing.

However, the 3-parameter laws are all very fine and quasi undistinguishable;   see Fig \ref{Fig5Plot36UEFA1112coco3flili}.
The high agreement between the resulting fits for the 53 UEFA members  is somewhat interesting. Note that the 3-parameter generalized exponential behavior "wins" much over the ZMP3 power law, from a $R^2$ rule point of view; see Table 2.

 Nevertheless, as for the FIFA country ranking\footnote{and for the UEFA team ranking \cite{MANUEFAteam}}, some marked deviation seems to occur  regularly.  In order to emphasize the occurrence of "regimes",    the  mere 2-parameter exponential and the power law fits are also presented in Fig. \ref{Fig6Plot8UEFA4yrnkcountries} for the top 20 "countries". Somewhat surprisingly the $R^2$ value of the exponential  fit slightly decreases ($\sim 0.982$), while the  $R^2$ value of the power law fit  reasonably increases  to $\sim 0.834$.  Observe that  the numerical parameter of the exponential is $ 1/0.08\simeq 12$. In fact, data steps can be read throughout the UEFA tables, near multiples of 6.  On one hand, it  can be claimed that such statistical results depend much on the finiteness of the data; thus, free to take these "small sample statistics" results as trends, rather than  securely established distributions.  On the other hand, they also point to inner structures.

   \section{On Rank-Rank correlation Laws}\label{rankrank}
  
 In order to observe whether there is or not some correlation between the two ranking schemes, one may  calculate the Kendall's $\tau$ rank measure  \cite{Kendalltau}. As done above, the FIFA all country ranking is projected onto the UEFA country set, and the ranking $r_{F,U}$ redefined  as $r$ ranging from 1 to 53.  The latter can be compared   to  the  mean ranking $r_{Uc}$ (and also to 
  the  point ranking $r_{Up}$, not done here). The Kendall's $\tau$ measure       \cite{Kendalltau} compares the number of concordant pairs $p$ and  non-concordant pairs $q$ through
  
 \begin{equation}\label{taueq}
\tau = \; \frac{p-q}{p+q}.
 \end{equation}
 Of course, $p+q= N(N-1)/2$, where $N$ is the number of  "agents" in the two  (necessarily equal size) sets.
  A website  \cite{Ktauweb} allows its immediate calculation. 
  
  It is found that $p =1100$ and $q = 278$, whence $\tau=0.5965$.  

Under the null hypothesis of independence of  the rank sets,  the sampling would   have an expected value $\tau = 0$. For large  samples, it is common to use an approximation to the normal distribution, with mean zero and variance, in order to emphasize the coefficient $\tau$  significance, through calculating
 \begin{equation}\label{tauvar}
 Z=\frac{\tau}{\sigma_{\tau}}\;\equiv   \frac{\tau}{\sqrt{\frac{2(2N+5)}{9N(N-1)}}}. 
     \end{equation}
    Since $N=53$,  and  $\sigma^2_{\tau}= 0.00895$ one  has $Z=6.3057 $, thereby indicating a  large correlation between the two sets.  
 
 Finally, a word on the  coefficient distributions is in order, before a specific discussion on country teams. The skewness is positive, in both cases,  necessarily here, see Table 1.  For such right skewed distributions,   most  data values are concentrated on left of the mean, with extreme values to the right.  Pearson's median (or second skewness coefficient) defined by $3(\mu-m)\sigma$ given in Table 1  confirms that most of the "area" is below the mean with  many countries having a high rank and low FIFA or UEFA coefficient value. On the other hand, the kurtosis is much below 3, indicating a platykurtic distribution, i.e. the data values are widely spread around the mean, - here with a "long" tail. In some sense, one obtains a picture of the Matthew effect: the winning countries are always  the same ones, and  stay more  at the top than others.

          \begin{table} \begin{center} 
\begin{tabular}[t]{cccccccccc} 
  \hline 
 
R$^2$  for &   ZMP3 &Pwco3&Stx3&Exp2 & fits  \\ 
\hline   	
 UEFA  &0.98& 0.99& 0.99& 0.99&   \\
\hline   	
FIFA(UEFA) &0.956& 0.973& 0.972& 0.968&   \\
\hline   	
FIFA(all) &0.962& 0.984& 0.981& 0.967&   \\
\hline  \end{tabular} 
   \caption{  Numerical  values of   the  regression coefficient $R^2$ for   fits  (see figures) with various empirical laws }\label{TableA1FIFA}
\end{center} \end{table}

\begin{table} \begin{center} 
\begin{tabular}[t]{cccccccccc} 
  \hline
rank	&	FIFA	&	UEFA	&	UEFA	&	FIFA	\\
	&	(UEFA)	&	(mean)	&	(points)	&	all	\\
\hline  
1	&	ESP	&	ESP	&	ESP	&	ESP	\\
2	&	GER	&	ENG	&	ENG	&	GER	\\
3	&	ENG	&	GER	&	GER	&	ENG	\\
4	&	POR	&	ITA	&	NED	&	POR	\\
5	&	ITA	&	POR	&	POR	&	URU	\\
6	&	NED	&	NED	&	ITA	&	ITA	\\
7	&	CRO	&	FRA	&	FRA	&	ARG	\\
8	&	DEN	&	RUS	&	BEL	&	NED	\\
9	&	GRE	&	BEL	&	RUS	&	CRO	\\
10	&	RUS	&	UKR	&	CYP	&	DEN	\\
11	&	FRA	&	GRE	&	UKR	&	GRE	\\
12	&	SWE	&	CYP	&	GRE	&	BRA	\\
 \hline
\end{tabular} 
   \caption{Top  twelve  soccer   countries in  Sept. 2012   ranked   according to different  coefficients   }\label{Tabletopcountries}
\end{center} \end{table} 

   \section{Discussion}\label{discussion}
  
   From a pragmatic point of view, it might be worthwhile to list the few top  countries according to their rank in Sep. 2012; see Table 3. The evolution of the country ranking in both schemes is outside the scope of the present study, concerned with  finding empirical laws (as discussed above) and rank-rank comparison, see below. Nevertheless, some  behavior of the dynamics of evolution is  manifested in the above data analysis and  through the finding of "regimes".  Note that along the above assertion,  some rational explanation of escaping from obeying exclusively either the power or the simple exponential laws can be imagined. This explanation also points implicitly to the classifying  groups, and their size,  imposed by the UEFA hierarchy in specific competitions.  
  
 In brief, England [ENG], Germany [GER]  and  Spain  [ESP] come  first. Next, [ITA], [NED], and [POR]  follow the three top  European nations\footnote{ After the  2013 all-German Champions League Final on May 25, 2013, between Borussia Dortmund and  Bayern M\"{u}nchen,  the position of Germany is  necessarily strengthened in the ranking, shifting it up to the level of Spain.
 
  Due to Paris Saint Germain  achievements,    this club has substituted Twente Enschede from the Netherlands in the team ranking, - therefore  dragging the Dutch score is as low as the Russian one, and pushing up France above these last two "Members".  The position of Italy  is rather stiff.
  
A  short comment  about  the ranking of  other specific countries,  under the logics of the presented comparison, seems in order here.. Note that the positions of Ukraine and Russia have become stiff or unchanged.  However, Poland and Bulgaria seem to loose systematically, - in terms of  football rankings, their reputation,   although  the two associations  (countries)  got EU membership after the great political turnover at the edge of the 1980s and the 1990s, - in contrast with Russia and Ukraine, both countries being still out of EU membership.
  Belgium,  being "exceptional", implies a comment in the text. }. There is a mild dissimilarity only between the two types of rankings. This is  expected from a sport fan point of view. Nevertheless, the concordance seems somewhat  puzzling because, on one hand the measures wear on different competitions, and on the other hand, the "European   club"  teams are far from being  composed of their citizens only. This remarks is substantiated by the case of Belgium, for which the national squad is quasi entirely made of players belonging to clubs outside Belgium, - the national squad having done rather well in the last few years, but the Belgian (club) teams made of diverse citizenship players are doing rather poorly in UEFA competitions.

 As of  now,  something seems to be clear  for  the UEFA (indirect) classification of countries: namely,  the majority rule of introducing (club) teams in the ranking by (national-like) UEFA  Association Members  goes in its first footing  
   scheme via a Fibonacci rule  or Pascal-like triangle.  
  It may also be viewed as a saturation effect in a finite system, i.e. the number of "valuable clubs', - together with the finite size of  the number of available days  for a one year competition. Bear in mind, however, that the Fibonacci rule is a signature of deterministic chaos to which no single scaling law applies at a reasonable length. In addition, recall that the Fibonacci ordering is characteristic of tree or plant  branchesÕ ramification to mention but two morphogenetic examples. On the other hand, it discloses the so called golden  division (or, golden ratio) of the whole, thus, a fairly rational principle of partitioning a whole into smaller pieces.  
  
  Those  inner structures might be mathematically further studied. Some hint for further investigation arises through some analogy. 
  These structures qualitatively resemble those of polycrystalline type materials
\cite{ADAM}. Namely, if a polycrystal had an intrinsically predetermined initial structure composed of a certain number of bigger-than-others grains, which is often the case, such grains would then ultimately survive following  (e.g. temperature) cycling \cite{hannay1992terrace,guillaume2005optimization} giving the largest contribution to the  final structure. It is because
such dominating (in size) grains suffer from more relaxed surface-tension conditions than their smaller  neighbors. Intriguingly, the  resulting grain distributions are found to be squeezed, as is also the Stx3  form used
for appropriate fitting, Eq. (\ref{Stxeq3}). They can be found obeying  the Weibull distributions \cite{ADAM}, with a scaling factor ($b$ therein), becoming time-independent in case when the  final structure
is formed. Thus, the structures pre-discovered in Fig.  \ref{Fig4Plot3FIFA0912regUEFAlili}  can qualitatively reflect a hierarchy
of grains in a polycrystal, grouping them in size-dependent pools. Note, however, that the
process of long-time polycrystal formation undergoes better statistics \cite{ADAMepl} than the
one presented here above, more considered as a trend in the present study.

  For the "countries" as for the team, the inner structure is likely tied to the pool, first rounds, then  to the  "direct elimination" tournament-like process. The no-power law regime seems therefore to indicate that indeed a tree rather than a network structure is at hand. Studies of round-robin tournaments with a large number of competitors/agents seem to be of interest for confirming, or not, the present findings.

 \section{Conclusions}\label{conclusions} 

 In conclusion, it has been searched whether two measuring schemes leading to ranking corroborate each other, although the schemes are geared toward different purposes. Moreover, it was searched whether the ranking leads to simple but empirical laws. 
The case of  European countries ranked by UEFA and FIFA from  different soccer  competitions was taken for  illustration.    It is found that the power law form  and subsequently   the rank size rule obviously  do  not appear  to be the best  simple description contrary to many modern expectations. In fact, it is found that the measuring rules lead to some inner structures, in both cases. These structures are proposed to originate from the types of competition and point attribution rules intrinsic to this complex system, i.e. a tree-like structure in contrast to a network structure. The latter is likely more appropriate in round-robin competitions.  {\it In fine},  Arrow's  impossibility theorem is also illustrated, since  the "ranked preferences" of teams are not univocal,  due to not  meeting a set of  "criteria" with three or more discrete options to choose from.
 
\vskip0.3cm     \begin{flushleft}
{\bf Acknowledgments} 
\end{flushleft}

MA and NKV  acknowledge some support through the project 'Evolution spatiale et temporelle d'infrastructures r\'egionales  et \' economiques en Bulgarie et en F{\'e}d{\'e}ration Wallonie-Bruxelles', within the   intergovernemental agreement for cooperation between the Republic of  Bulgaria and  the  Communaut\' e  Fran\c{c}aise de Belgique. Moreover, this work has been performed in the framework of COST Action IS1104  "The EU in the new economic complex geography: models, tools and policy evaluation".

\vskip0.3cm

 \begin{appendix} {\bf Appendix  A. All FIFA countries}
 \vskip0.3cm     
 For completeness, let this Appendix contain a display and short analysis of the FIFA ranking for the 206 countries which level was recorded in 2012. Note that  beside the UEFA  
San Marino,  three countries, i.e. Bhutan, Montserrat,  and 
Turk  and  Caicos Islands, had  a coefficient equal to 0.
 
 Table A1: numerical  values of parameter fits and  the corresponding regression coefficient $R^2$ for all FIFA countries in  Sept. 2012; see Fig1A for illustration

\begin{figure}
\centering
 \includegraphics [height=16.5cm,width=14.5cm]{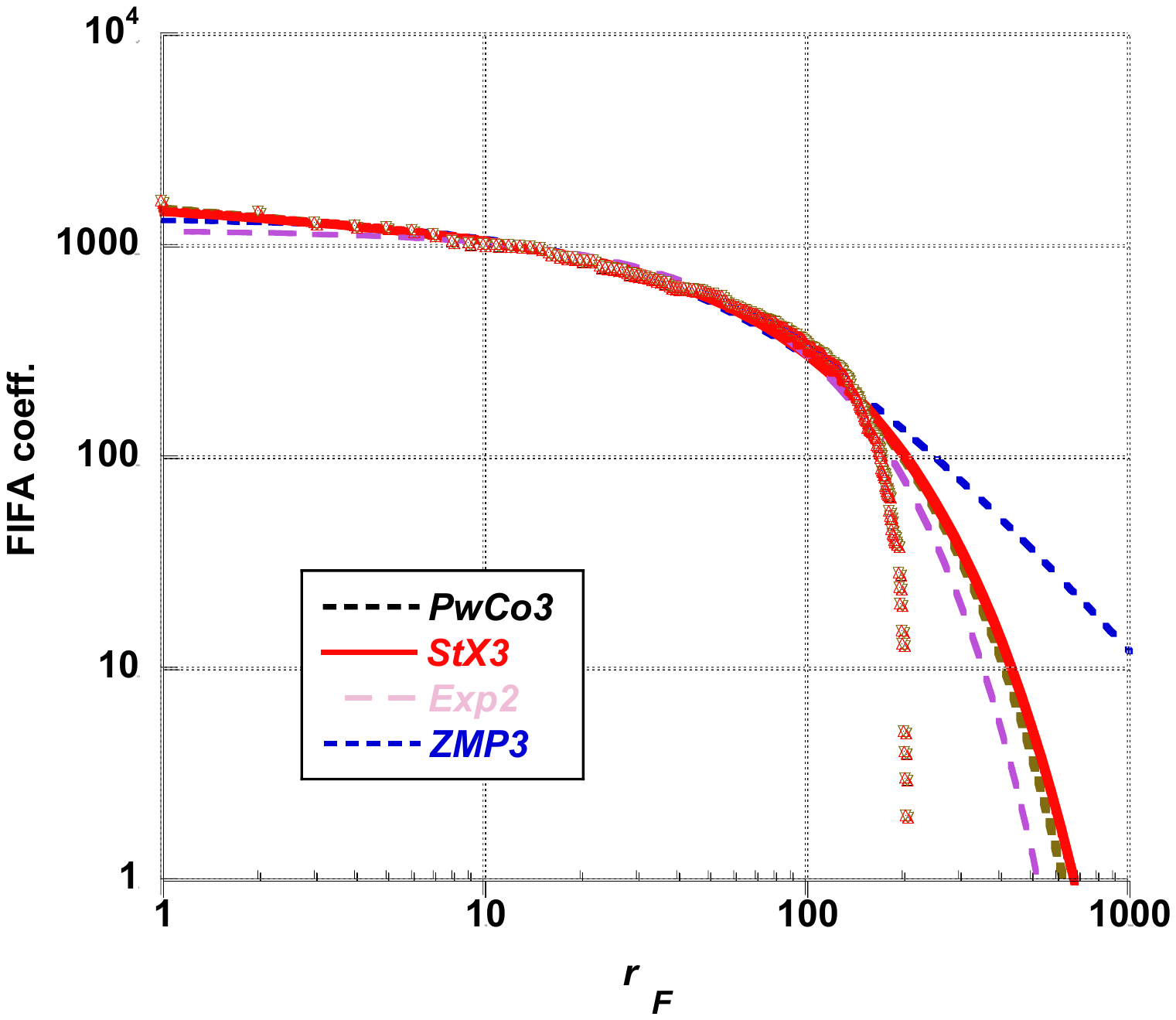}
\caption   {   Empirical relationships between  the  FIFA countries rank  in Sept. 2012;  
the corresponding regression coefficient $R^2$ are given in Table A1 } 
 \label{Fig1AFIFAallcountries} 
\end{figure}

  A few possible and simple empirical relationships, as proposed in the main text,  between  the  FIFA country coefficient as ranked  in Sept. 2012 are shown in Fig. \ref{Fig1AFIFAallcountries} on    log-log plots. Four fits are displayed;  
the corresponding regression coefficients $R^2$ are given in Table 2.  Except for the Exp2 fit which seems to miss the low ranked data,  the 3-parameter empirical laws seem to be equivalently precise. 

\begin{figure}
\centering
 \includegraphics [height=16.5cm,width=14.5cm]{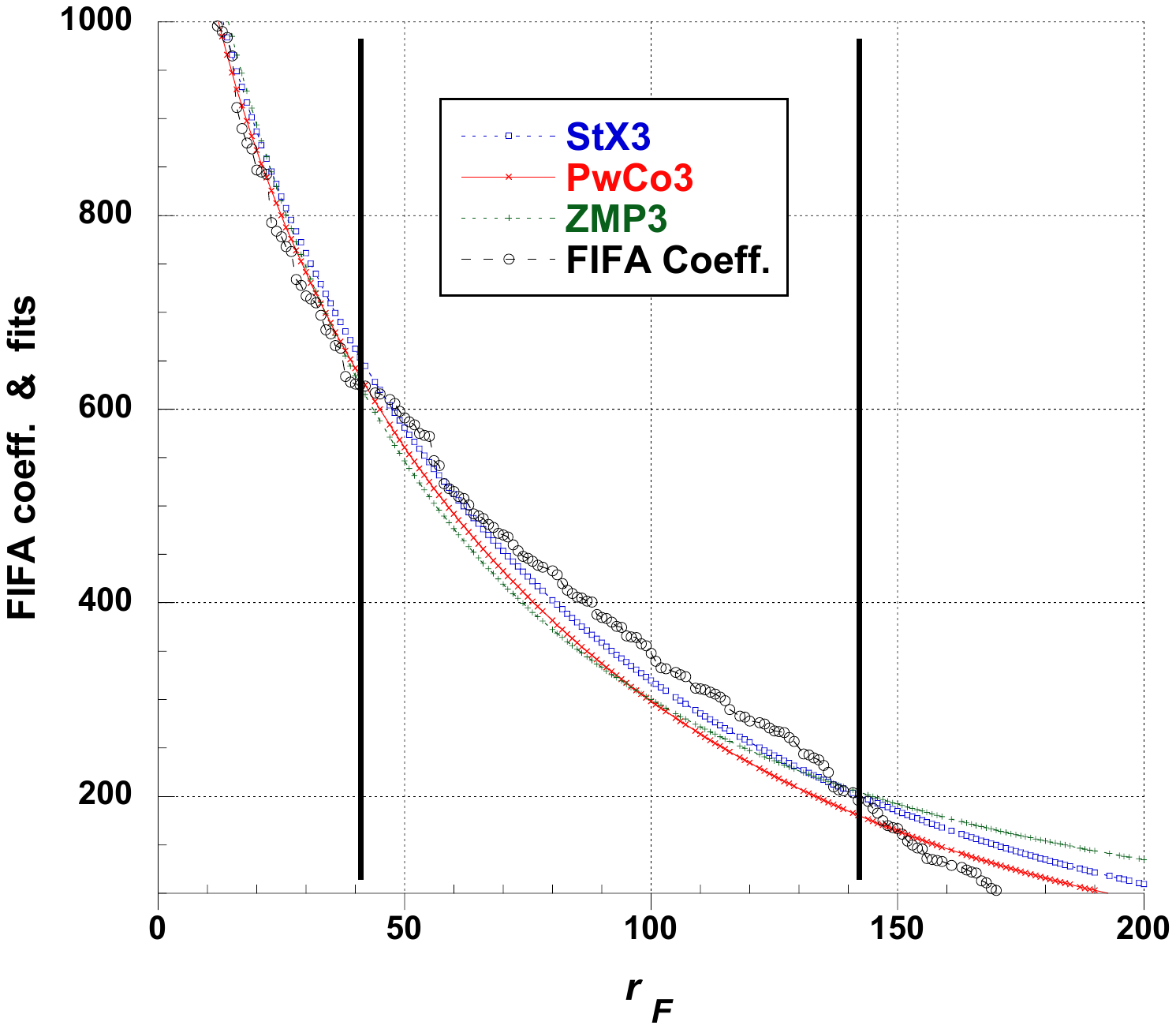}
\caption   {   Indication of rank regimes in the  FIFA countries in Sept. 2012;  
fits; dark vertical lines suggest regime ranges, i.e. when data crosses fits  } 
 \label{Fig2APlot44liliintersect}
\end{figure}

Nevertheless some marked deviation occurs for $r\ge  100$. A more detailed examination of the  the difference between the data and the fits indicate large regions where the difference is either positive or negative.  Intersection points  depend on the empirical law, but occur in  a close $r$ range.  The sign of the difference defines  regimes limited either by $r\le45$ and $r\ge145$ for  Stx3  and Pwco3 or  by $r\le40$ and $r\ge140$ for  ZMP3.  This is illustrated in Fig.\ref{Fig2APlot44liliintersect}. In some colloquial way, one can thus  point to three categories of countries: (i) the top ones, (i)) the "soft belly", and (iii) the "small" ones.

  \end{appendix}
\vskip 0.5cm

 \begin{appendix} {\bf Appendix  B. Correlation between UEFA Points and UEFA Average Coefficients }
 \vskip 0.5cm

     \begin{figure}
\centering
 \includegraphics [height=16.5cm,width=14.5cm]{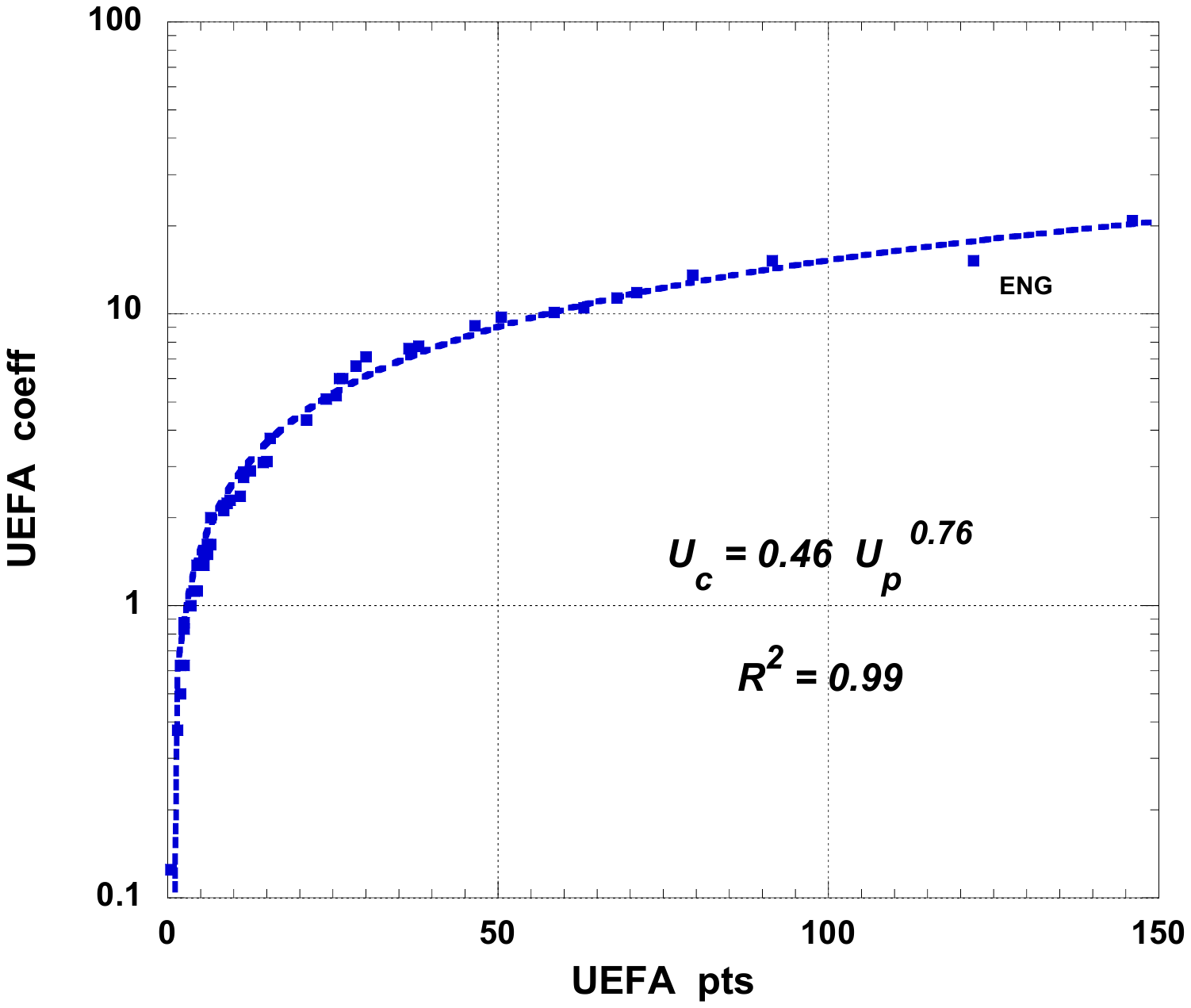}
\caption   {  Correlation between UEFA points and UEFA coefficients in Sept 2012:  a power law fit  } 
 \label{Fig1BPlot64upuc}
\end{figure}

The correlation between reordered UEFA points ($U_p$) and UEFA coefficients ($U_c$)  in Sept 2012
\begin{equation}\label{U_cU_p}
U_c = U_p /N
\end{equation}
is shown in    Fig.\ref{Fig1BPlot64upuc}.  There is hardly an explanation for this remarkable fit, nor any reason why ENG is outside the fit. 

The number of concordance pairs $p$ and  non-concordance pairs $q$  of the $r_{Uc}$ and  $r_{Up}$  leads to $\tau= 0.9376$ from Eq.(\ref{taueq}); $p-q$= 1292 and $p+q$= 1378, or $p$=1335 and $q$= 43.   
This large value of $\tau$ indicates that the two measures strongly agree, thus are rather equivalent,
on the "evaluation of the countries".

Moreover from Eq.(\ref{tauvar}),  since $N=53$,  whence  $\sigma_{\tau}= 0.0946$, one  has $Z=9.911$, -  quite large  to reject the null  hypothesis, indicating a high correlation between the two sets.
   \end{appendix}

   \end{document}